\documentclass[nofootinbib,twocolumn,preprintnumbers,superscriptaddress]{revtex4}
\usepackage{amsmath}
\usepackage{amssymb}
\usepackage{graphicx}
\usepackage{hyperref}
\usepackage{xspace}

\usepackage{booktabs}
\AtBeginDocument{
\heavyrulewidth=.08em
\lightrulewidth=.05em
\cmidrulewidth=.03em
\belowrulesep=.65ex
\belowbottomsep=0pt
\aboverulesep=.4ex
\abovetopsep=0pt
\cmidrulesep=\doublerulesep
\cmidrulekern=.5em
\defaultaddspace=.5em
}

\newcommand{\GeV}{\;\mathrm{GeV}}

\newcommand{\POWHEGBOX}{\texttt{POWHEG-BOX}\xspace}
\newcommand{\POWHEG}{\texttt{POWHEG}\xspace}
\newcommand{\PYTHIA}{\texttt{PYTHIA}\xspace}

\usepackage{xcolor}
\definecolor{light-gray}{gray}{0.8}

\begin{document}

\title{Fully differential VBF Higgs production at NNLO}

\preprint{CERN-PH-TH/2015-127}
\preprint{OUTP-15-12P}

\newcommand{\CERNaff}{CERN, PH-TH, CH-1211 Geneva 23, Switzerland}
\newcommand{\DIDaff}{Universit\'e Paris Diderot, Paris, France}
\newcommand{\CNRSaff}{CNRS, UMR 7589, LPTHE, F-75005, Paris, France}
\newcommand{\SORBaff}{Sorbonne Universit\'es, UPMC Univ Paris 06, UMR
  7589, LPTHE, F-75005, Paris, France}
\newcommand{\OXFaff}{Rudolf Peierls Centre for Theoretical Physics,
  1 Keble Road, University of Oxford, UK}

\author{Matteo Cacciari}
\affiliation{\DIDaff}
\affiliation{\SORBaff}
\affiliation{\CNRSaff}
\affiliation{\CERNaff}
\author{Fr\'ed\'eric A. Dreyer}
\affiliation{\SORBaff}
\affiliation{\CNRSaff}
\affiliation{\CERNaff}
\author{Alexander Karlberg}
\affiliation{\OXFaff}
\author{Gavin P. Salam}
\altaffiliation{On leave from \CNRSaff}
\affiliation{\CERNaff}
\author{Giulia Zanderighi}
\affiliation{\CERNaff}
\affiliation{\OXFaff}

\begin{abstract}
  We calculate the fully differential next-to-next-to-leading-order
  (NNLO) corrections to vector-boson fusion (VBF) Higgs production at
  proton colliders, in the limit in which there is no cross-talk
  between the hadronic systems associated with the two protons.
  We achieve this using a new ``projection-to-Born'' method that
  combines an inclusive NNLO calculation in the structure-function
  approach and a suitably factorised next-to-leading-order (NLO) VBF Higgs plus 3-jet
  calculation, using appropriate Higgs plus 2-parton counterevents.
  An earlier calculation of the fully inclusive cross section had
  found small NNLO corrections, at the $1\%$ level.
  In contrast, the cross section after typical experimental VBF cuts
  receives NNLO contributions of about 4\%, while differential
  distributions show corrections of up to 6-7\% for some standard
  observables.
  The corrections are often outside the NLO scale-uncertainty band. 
\end{abstract}


\maketitle

Following the discovery in 2012 of the Higgs
boson~\cite{Aad:2012tfa,Chatrchyan:2012ufa}, one of the main tasks for
particle physics today is the accurate determination of its properties
and couplings.
For the coming decade at least, these studies will take place at
CERN's Large Hadron Collider (LHC).

The most relevant production channels for the Higgs boson at the LHC
are gluon fusion (ggH), vector-boson fusion (VBFH), production in
association with a vector boson (VH) and with a top-quark pair
(ttH)~\cite{Dittmaier:2011ti}.
VBFH is special for a number of
reasons~\cite{Dittmaier:2012vm,Heinemeyer:2013tqa}: it has the largest
cross section of the processes that involves tree-level production of
the Higgs boson (and is second largest among all processes); it has a
distinctive signature of two forward jets, which makes it possible to
tag the events and so identify Higgs decays that normally have large
backgrounds, e.g.\ $H \to \tau^+ \tau^-$; the Higgs transverse
momentum is non-zero even at lowest order, which facilitates searches
for invisible decay
modes~\cite{ATLAS-VBFH-invisible,CMS-VBFH-invisible}; and it also
brings particular sensitivity to the charge-parity properties of the
Higgs boson, and non-standard Higgs interactions, through the angular
correlations of the forward jets~\cite{Plehn:2001nj}.

Given the key role of VBF Higgs-boson production at the LHC, it is of
paramount importance to have a precise prediction for its production.
The total cross section was calculated to NNLO in
Refs.~\cite{Bolzoni:2010xr,Bolzoni:2011cu} using the
structure-function approach~\cite{Han:1992hr}, showing small
corrections relative to the NLO and tiny renormalisation and
factorisation scale uncertainties, well below $1\%$.
However experimental measurements are necessarily restricted to a
subset of phase space. In particular, because of their use of
transverse-momentum cuts on the forward tagging jets, one might
imagine that there are important NNLO corrections, associated with
those jet cuts, that would not be seen in a fully inclusive
calculation.
Currently, the fully differential VBFH cross section is known only to
NLO~\cite{Figy:2003nv}.
It appears to have small scale uncertainties.

\begin{figure*}
  \centering
  \includegraphics[width=\textwidth]{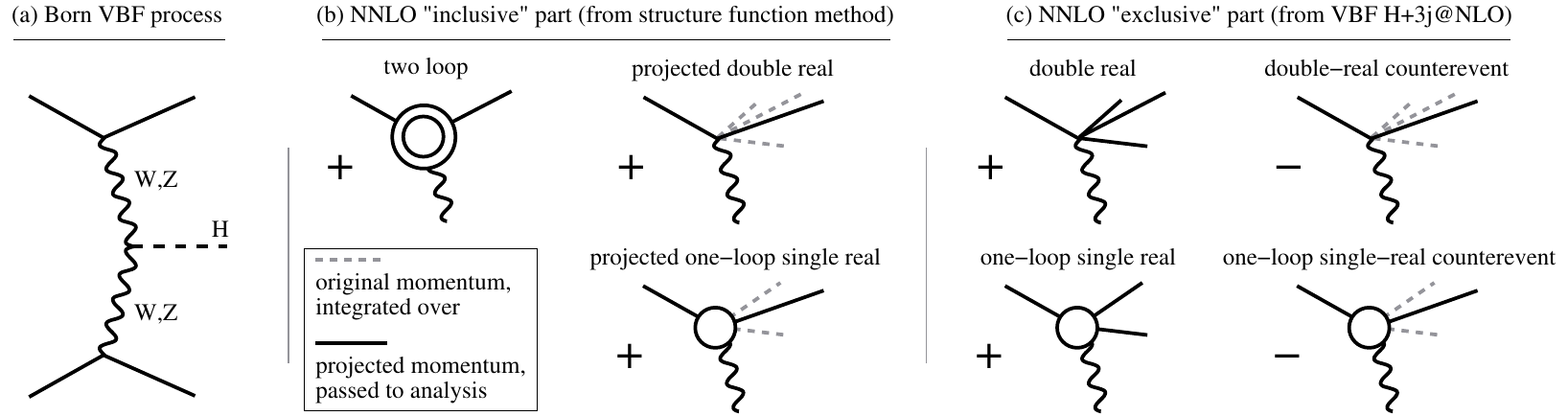}
  \caption{(a) Illustration of the Born VBFH process. 
    (b) NNLO corrections to the upper sector of the VBF process, from the
    ``inclusive'' part of our calculation.
    (c) Corresponding ``exclusive'' part.
    The double-real and one-loop single-real counterevents in the
    exclusive part cancel the projected double-real and one-loop
    single-real contributions in the inclusive part.
    In the ``projected'' and ``counterevent'' contributions, the dashed
    lines corresponds to the full set of parton momenta that are
    integrated over (for the structure functions, this integral is
    implicit in the derivation of the coefficient functions), while the
    solid lines correspond to the partons that are left over after
    projection to Born-like kinematics and then passed to the analysis. 
    The projection does not change the direction of initial partons
    and so the corresponding incoming dashed lines are implicit.
  }
  \label{fig:ingredients}
\end{figure*}

One can think of the VBFH process, represented at the Born level in
Fig.~\ref{fig:ingredients}a, as involving two Deeply Inelastic
Scatterings (DIS), one for each of the incoming protons.
Each DIS process produces a vector boson, $W^\pm$ or $Z$, and the
fusion of the vector bosons produces a Higgs boson.
The structure-function approach~\cite{Han:1992hr} used for the NNLO
total cross section~\cite{Bolzoni:2010xr,Bolzoni:2011cu} assumes that
the upper and lower hadronic sectors factorise from each other, i.e.\
that there is no cross-talk between them.
Factorisation is believed to be accurate to better than
$1\%$ in the experimentally relevant kinematic
region~\cite{Ciccolini:2007ec,Andersen:2007mp,Bolzoni:2011cu,Zaro:2013twa}.%
\footnote{The factorisation of the two sectors is exact if one
  imagines two copies of QCD, QCD$_1$ and QCD$_2$, respectively for
  the upper and lower sectors, where each of the two QCD copies
  interacts with the electroweak (EW) sector, but not with the other
  QCD copy.
  This observation could be exploited, for example, in automated
  calculations and for determining corrections beyond the factorised
  approximation.}

The reason that the structure function approach does not provide a
fully differential cross section is related to the fact that the DIS
coefficient functions used in the calculation implicitly integrate
over hadronic final states.
In this letter, we introduce a new ``projection-to-Born'' approach to
eliminate this limitation, and thus provide the first fully
differential NNLO calculation of VBF Higgs production in the
factorised approximation.

Let us start by recalling that the cross section in
the structure-function approach 
is expressed~\cite{Han:1992hr,Bolzoni:2010xr,Bolzoni:2011cu} as
a sum of terms involving products of structure functions, e.g.
$F_2(x_1, Q_1^2) F_2(x_2, Q_2^2)$, where $Q_i^2 = -q_i^2 > 0$ is given
in terms of the 4-momentum $q_i$ of the (outgoing) exchanged vector
boson $i$.
The $x_i$ values are fixed by the relation $x_i = - Q_i^2/(2
P_i.q_i)$, where $P_i$ is the momentum of proton $i$.
To obtain the total cross section, one integrates over all $q_1$,
$q_2$ that can lead to the production of a Higgs boson.
If the underlying upper (lower) scattering is Born-like, $\text{quark}
\to \text{quark} + V$, then it is straightforward to show that
knowledge of the vector-boson momentum $q_1$ ($q_2$) uniquely
determines the momenta of both the incoming and outgoing (on-shell)
quarks,
\begin{equation}
  \label{eq:kinematics}
  p_{\text{in},i} = x_i P_i,\qquad p_{\text{out},i} = x_i P_i - q_i\,.
\end{equation}

We exploit this feature in order to assemble a full calculation from
two separate ingredients.
For the first one, the ``inclusive'' ingredient, we remain within the
structure function approach, and for each set of $q_1$ and $q_2$ use
Eq.~(\ref{eq:kinematics}) to assign VBF Born-like kinematics to the
upper and lower sectors.
This is represented in Fig.~\ref{fig:ingredients}b (showing, for
brevity, just the upper sector): for the two-loop contribution, the
Born kinematics that we assign corresponds to that of the actual
diagrams;
for the tree-level double-real and one-loop single-real diagrams, it
corresponds to a projection from the true kinematics ($2\to H+n$ for
$n=3,4$) down to the Born kinematics ($2\to H+2$).
The projected momenta are used to obtain the ``inclusive''
contribution to differential cross sections.
Note that the Higgs momentum is unaffected by the projection.

Our second, ``exclusive'', ingredient starts from the NLO fully
differential calculation of vector-boson fusion Higgs production with
three jets~\cite{Figy:2007kv,Jager:2014vna}, as obtained in a
factorised approximation, i.e.\ where there is no cross-talk between
upper and lower sectors.\footnote{The NLO calculation without this
  approximation is given in Ref.~\protect\cite{Campanario:2013fsa}.}
Thus each parton can be uniquely assigned to one of the upper or lower
sectors and the two vector-boson momenta can be unambiguously
determined.
For each event in a Monte Carlo integration over phase space, with
weight $w$, we add a counterevent, with weight $-w$, to which we
assign projected Born VBF kinematics based on the vector-boson momenta
and Eq.~(\ref{eq:kinematics}).
This is illustrated in Fig.~\ref{fig:ingredients}c.
From the original events, we thus obtain the full momentum structure
for tree-level double-real and one-loop single-real contributions.
Meanwhile, after integration over phase space, the counterevents
exactly cancel the projected tree-level double-real and one-loop
single-real contributions from the inclusive part of the calculation.
Thus the sum of the inclusive and exclusive parts gives the complete
differential NNLO VBFH result.%
\footnote{Our approach can be contrasted with the differential NNLO
  structure-function type calculation for single-top
  production~\cite{Brucherseifer:2014ama} in that we do not need any
  fully differential ingredients at NNLO.}

For the implementation of the inclusive part of the calculation, we
have taken the phase space from \POWHEG's Higgs plus two-jet VBF
calculation~\cite{Nason:2009ai}, while the matrix element has been
coded with structure functions evaluated using parametrised
versions~\cite{vanNeerven:1999ca,vanNeerven:2000uj} of the NNLO DIS
coefficient
functions~\cite{vanNeerven:1991nn,Zijlstra:1992qd,Zijlstra:1992kj}
integrated with \texttt{HOPPET} v1.1.5~\cite{Salam:2008qg}.
We have tested our implementation against the results of one of the
codes used in Ref.~\cite{Bolzoni:2010xr,Bolzoni:2011cu} and found
agreement, both for the structure functions and the final cross
sections.
We have also checked that switching to the exact DIS coefficient
functions has a negligible impact.
A further successful comparison of the evaluation of structure
functions was made against
\texttt{APFEL}~v.2.4.1\cite{Bertone:2013vaa}.

For the exclusive part of the calculation, as a starting point we took
the NLO (i.e.\ fixed-order, but not parton-shower) part of the \POWHEG
$H$+3-jet VBF code~\cite{Jager:2014vna}, itself based on the
calculation of Ref.~\cite{Figy:2007kv}, with tree-level matrix
elements from MadGraph~4~\cite{Alwall:2007st}.
This code already uses a factorised approximation for the matrix
element, however for a given phase-space point it sums over
matrix-element weights for the assignments of partons to upper and
lower sectors.
We therefore re-engineered the code so that for each set of 4-momenta,
weights are decomposed into the contributions for each of the
different possible sets of assignments of partons to the two sectors.
For every element of this decomposition it is then possible to
unambiguously obtain the vector-boson momenta and so correctly
generate a counterevent.
The \POWHEGBOX's~\cite{Nason:2004rx,Alioli:2010xd} ``tagging''
facility was particularly useful in this respect, notably for the NLO
subtraction terms.
To check the correctness of the assignment to sectors, we verified
that as the rapidity separation between the two leading jets
increases, there was a decreasing relative fraction of the cross
section for which partons assigned to the upper (lower) sector
were found in the rapidity region associated with the lower (upper)
leading jet.
We also tested that the sum of inclusive and exclusive contributions
at NLO agrees with the \POWHEG NLO implementation of the VBF $H$+2-jet
process.

To investigate the phenomenological consequences of the NNLO
corrections, we study 13 TeV proton-proton collisions.
We use a diagonal CKM matrix, full Breit-Wigners for the $W$, $Z$ and
the narrow-width approximation for the Higgs boson.
We take NNPDF 3.0 parton distribution functions at NNLO with
$\alpha_s(M_Z) = 0.118$
(\texttt{NNPDF30\_nnlo\_as\_0118})~\cite{Ball:2014uwa}, also for our
LO and NLO results.
We have five light flavours and ignore contributions with top-quarks
in the final state or internal lines.
We set the Higgs mass to $M_H = 125\GeV$, compatible with the
experimentally measured value~\cite{Aad:2015zhl}.  Electroweak
parameters are set according to known experimental values and
tree-level electroweak relations. As inputs we use $M_W = 80.398\GeV$,
$M_Z = 91.1876\GeV$ and $G_F = 1.16637\times 10^{-5} \GeV^{-1}$. For
the widths of the vector bosons we use $\Gamma_W = 2.141 \GeV $ and
$\Gamma_Z = 2.4952 \GeV$.

Some care is needed with the renormalisation and factorisation scale
choice.
A natural option would be to use $Q_1$ and $Q_2$ as our central values
for the upper and lower sectors, respectively.
While this is straightforward in the inclusive code, in the exclusive
code we had the limitation that the underlying \POWHEGBOX code can
presently only easily assign a single scale (or set of scales) to a
given event.
However, for each \POWHEG phase-space point, we have multiple
upper/lower classifications of the partons, leading to several
$\{Q_1,Q_2\}$ pairs for each event.
Thus the use of $Q_1$ and $Q_2$ would require some further degree of
modification of the \POWHEGBOX, which we leave to future work.
We instead choose a central scale that depends on the Higgs transverse
momentum $p_{t,H}$:
\begin{equation}
  \label{eq:scale}
  \mu_0^2(p_{t,H}) = \frac{M_H}{2} \sqrt{\left(\frac{M_H}{2}\right)^2 +
        p_{t,H}^2}\,.
\end{equation}
This choice of $\mu_0$ is usually close to
$\sqrt{Q_1 Q_2}$.
It represents a good compromise between satisfying the requirement of
a single scale for each event, while dynamically adapting to the
structure of the event.
In order to estimate missing higher-order uncertainties, we vary the
renormalisation and factorisation scales symmetrically (i.e. keeping
$\mu_R=\mu_F$) by a factor $2$ up and down around $\mu_0$.%
\footnote{We verified that an expanded scale variation, allowing
  $\mu_R \neq \mu_F$ with $\frac12 < \mu_R/\mu_F < 2$, led only to
  very small changes in the NNLO scale uncertainties for the VBF-cut
  cross section and the $p_{t,H}$ distribution.}

To pass our VBF selection cuts, events should have at least two jets
with transverse momentum $p_t > 25\GeV$; the two hardest (i.e.\
highest $p_t$) jets should have absolute rapidity $|y|<4.5$, be
separated by a rapidity $\Delta y_{j_1,j_2} > 4.5$, have a dijet
invariant mass $m_{j_1,j_2} > 600\GeV$ and be in opposite hemispheres
($y_{j_1} y_{j_2} < 0$).
Jets are defined using the anti-$k_t$
algorithm~\cite{Cacciari:2008gp}, as implemented in \texttt{FastJet
  v3.1.2}~\cite{Cacciari:2011ma}, with radius parameter $R=0.4$.

\begin{table}[t] 
  \centering
  \phantom{x}\medskip
  \begin{tabular}{lcccc}
    \toprule
    &&  $\sigma^\text{(no cuts)}$  [pb]  && $\sigma^\text{(VBF cuts)}$ [pb] \\
    \midrule
    LO      &&  $4.032\,^{+0.057}_{-0.069}$    &&  $0.957\,^{+0.066}_{-0.059}$\\[4pt]
    NLO     &&  $3.929\,^{+0.024}_{-0.023}$    &&  $0.876\,^{+0.008}_{-0.018}$\\[4pt]
    NNLO    &&  $3.888\,^{+0.016}_{-0.012}$    &&  $0.844\,^{+0.008}_{-0.008}$\\
    \bottomrule
  \end{tabular}
  \caption{Cross sections at LO, NLO and NNLO for VBF Higgs production,
    fully inclusively and with VBF cuts.
    The quoted uncertainties correspond to scale dependence, while
    statistical errors at NNLO are about $0.1\%$ with VBF cuts and much smaller 
    without. }
\label{tab:cross-sections}
\end{table}

Results are shown in table~\ref{tab:cross-sections} for the fully
inclusive cross section and with our VBF cuts.
One sees that the NNLO corrections modify the fully inclusive cross
section only at the percent level, which is compatible with the
findings of Ref.~\cite{Bolzoni:2010xr}.
However, after VBF cuts, the NNLO corrections are $3{-}4$ times
larger, reducing the cross section by $4\%$ relative to NLO.
The magnitude of the NNLO effects after cuts implies that it will be
essential to take them into account for future precision studies.
Note that in both the inclusive and VBF-cut cases, the NNLO
contributions are larger than would be expected from NLO scale
variation.

\begin{figure*}
  \centering
          \includegraphics[clip,height=0.45\textwidth,page=1,angle=0]{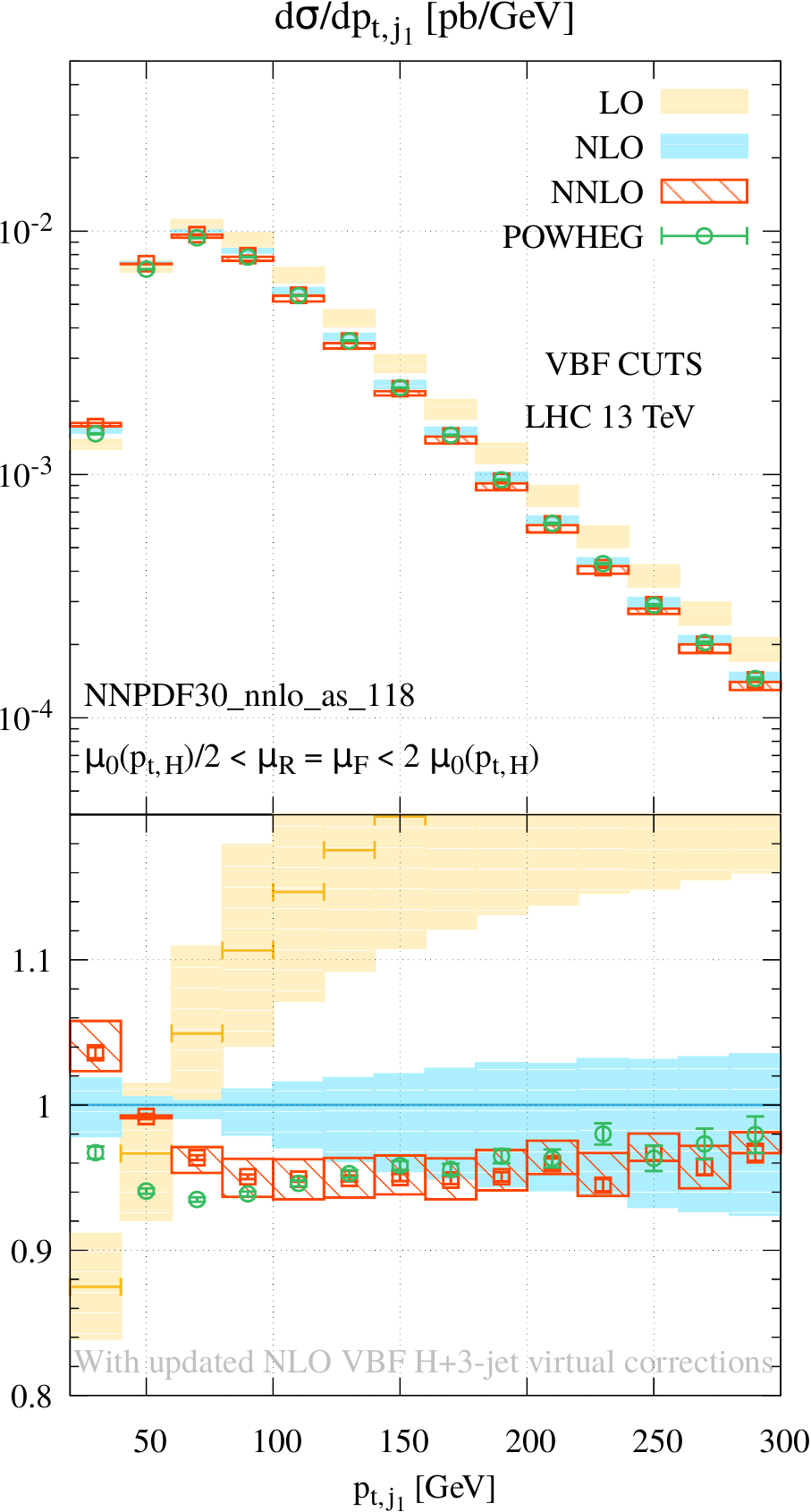}%
    \hfill\includegraphics[clip,height=0.45\textwidth,page=2,angle=0]{bugfix-crop.pdf}%
    \hfill\includegraphics[clip,height=0.45\textwidth,page=3,angle=0]{bugfix-crop.pdf}\hspace{0.8mm}%
    \hfill\includegraphics[clip,height=0.45\textwidth,page=4,angle=0]{bugfix-crop.pdf}%
  \caption{From left to right, differential cross sections for the transverse momentum
distributions  for the two leading jets,
$p_{t,j_1}$ and $p_{t,j_2}$, for the Higgs boson, $p_{t,H}$, and the distribution for the rapidity 
separation between the two leading jets, $\Delta y_{j_1,j_2}$.  }
  \label{fig:diff-cross-sections}
\end{figure*}

Differential cross sections are shown in
Fig.~\ref{fig:diff-cross-sections}, for events that pass the VBF cuts.
From left to right, the plot shows the transverse momentum
distributions for the two leading jets, $p_{t,j_1}$ and $p_{t,j_2}$,
for the Higgs boson, $p_{t,H}$, and the distribution for the rapidity
separation between the two leading jets, $\Delta y_{j_1,j_2}$.
The bands and the patterned boxes denote the scale uncertainties,
while the vertical error-bars denote the statistical uncertainty.
The effect of the NNLO corrections on the jets appears to be to reduce
their transverse momentum, leading to negative (positive) corrections
in regions of falling (rising) jet spectra.
One can see effects of up to $6{-}7\%$.
Turning to $p_{t,H}$, one might initially be surprised that such an
inclusive observable should also have substantial NNLO corrections, of
about $5\%$ for low and moderate $p_{t,H}$.
Our interpretation is that since NNLO effects redistribute jets from
higher to lower $p_t$'s (cf.\ the plots for $p_{t,j_1}$ and
$p_{t,j_2}$), they reduce the cross section for any observable defined
with VBF cuts.
As $p_{t,H}$ grows larger, the forward jets tend naturally to
get harder and so automatically pass the $p_t$ thresholds, reducing
the impact of NNLO terms. 

As observed above for the total cross section with VBF cuts, the NNLO
differential corrections are sizeable and often outside the
uncertainty band suggested by NLO scale variation.
One reason for this might be that NLO is the first order where the
non-inclusiveness of the jet definition matters, e.g.\ radiation
outside the cone modifies the cross section.
Thus NLO is, in effect, a leading-order calculation for the exclusive
corrections, with all associated limitations.

To further understand the size of the NNLO corrections, it is
instructive to examine a NLO plus parton shower (NLOPS) calculation,
since the parton shower will include some approximation of the NNLO
corrections.
For this purpose we have used the \POWHEG VBF $H$+2-jet
calculation~\cite{Nason:2009ai}, showered with \PYTHIA version 6.428
with the Perugia 2012 tune~\cite{Skands:2010ak}.
The \POWHEG part of this NLOPS calculation uses the same PDF, scale
choices and electroweak parameters as our full NNLO calculation. 
The NLOPS results are included in Fig.~\ref{fig:diff-cross-sections}, at
parton level, with multi-parton interactions (MPI) switched off.
They differ from the NLO by an amount that is of a similar order of
magnitude to the NNLO effects.
This lends support to our interpretation that final (and
initial)-state radiation from the hard partons is responsible for a
substantial part of the NNLO corrections.
However, while the NLOPS calculation reproduces the shape of the NNLO
corrections for some observables (especially $p_{t,H}$), there are
others for which this is not the case, the most
striking being perhaps $\Delta y_{j_1,j_2}$.
Parton shower effects were also studied in
Ref.~\cite{Frixione:2013mta}, using the MC@NLO
approach~\cite{Frixione:2002ik}.
Various parton showers differed there by up to about 10\%.

In addition to the NNLO contributions, precise phenomenological
studies require the inclusion of EW contributions and non-perturbative
hadronisation and MPI corrections.
The former are of the same order of magnitude as our NNLO
corrections~\cite{Ciccolini:2007ec}.
Using Pythia~6.428 and Pythia~8.185 we find that hadronisation
corrections are between $-2$ and $0\%$, while MPI brings up to $+5\%$
at low $p_t$'s.
The small hadronisation corrections appear to be due to a partial
cancellation between shifts in $p_t$ and rapidity.
We leave a combined study of all effects to future work.
The code for our calculation will also be made public.

With the calculation presented in this letter, differential VBF Higgs
production has been brought to the same NNLO level of accuracy that
has been available for some time now for the
ggH~\cite{Catani:2007vq,Anastasiou:2004xq} and
VH~\cite{Ferrera:2011bk} production channels.
This constitutes the first fully differential NNLO $2\to 3$
hadron-collider calculation, an advance made possible thanks to the
factorisable nature of the process.
The NNLO corrections are non-negligible, $4$--$7\%$, almost an order
of magnitude larger than the corrections to the inclusive cross
section.
Their size might even motivate a calculation one order higher, to
N$^3$LO, to match the precision achieved recently for the ggH total
cross section~\cite{Anastasiou:2015ema}.
With the new ``projection-to-Born'' approach introduced here, we
believe that this is within reach.
It would also be of interest to obtain NNLO plus parton shower
predictions, again matching the accuracy achieved recently in
ggH~\cite{Hamilton:2013fea,Hoche:2014dla}.

\textbf{Acknowledgments:} 
We wish to thank Marco Zaro for sharing his private code for the
inclusive NNLO calculation, which enabled valuable cross checks, and
for numerous useful discussions as well as comments on the manuscript.
We also thank Fabrizio Caola for his careful reading of the manuscript 
and valuable comments.
We are grateful to Paolo Nason for his suggestions concerning the
adaptation of the \POWHEG{} code, to Barbara J\"ager for helpful
exchanges about the VBF Higgs plus 3-jet process, to Andreas Vogt for
supplying us with the exact coefficient DIS coefficient functions and
to Valerio Bertone and Juan Rojo for discussions about APFEL.
We also benefitted from discussions with John Ellis, Zoltan Kunszt and
Fabio Maltoni.
FD is supported by the ILP LABEX (ANR-10-LABX-63) financed by French
state funds managed by the ANR within the Investissements d'Avenir
programme under reference ANR-11-IDEX-0004-02.
AK is supported by the British Science and Technology Facilities
Council and by the Buckee Scholarship at Merton College.
This work was supported also by ERC Consolidator Grant HICCUP and by
ERC Advanced Grant Higgs@LHC.
AK thanks CERN for hospitality while part of this work was performed.

For version 2 of this letter we are grateful to Juan Cruz-Martinez,
Thomas Gehrmann, Nigel Glover and Alexander Huss (CGGH), as well as
Barbara J\"ager, for bringing to our attention a bug and associated
fix in the virtual terms of the original NLO VBF Higgs plus 3-jet
calculation~\cite{Figy:2007kv} and its POWHEG
implementation~\cite{Jager:2014vna} and to CGGH for detailed
comparisons with their results~\cite{Cruz-Martinez:2018rod} prior to
publication. Their results and those from our version 2 are in good
agreement.
%


\end{document}